\documentclass[12pt,preprint]{aastex}
\usepackage{verbatim}
\usepackage{natbib}

\shorttitle{A New Asteroseismic Probe of the Helium Core Flash}
\shortauthors{Bildsten, Paxton, Moore and Macias}

\begin{document}

\newcommand{\Msun}{\mathrm{M_{\odot}}}
\newcommand{\Lsun}{\mathrm{L_{\odot}}}
\newcommand{\Rsun}{\mathrm{R_{\odot}}}
\newcommand{\Teff}{\mathrm{T_{eff}}}
\newcommand{\logg}{\mathrm{log(g)}}
\newcommand{\logTeff}{log~\Teff}
\newcommand{\MESAstar}{\texttt{MESA star}}
\newcommand{\MESA}{\texttt{MESA}}

\title{Acoustic Signatures of the Helium Core Flash}

\author{Lars Bildsten, Bill Paxton, Kevin Moore, \& Phillip J. Macias}
\affil{Kavli Institute for Theoretical Physics and Department of Physics\\
Kohn Hall, University of California, Santa Barbara, CA 93106\\}

\begin{abstract} 

All evolved stars with masses $M \lesssim 2 M_\odot$ undergo an
initiating off-center helium core flash in their $M_c\approx
0.48M_\odot$ He core as they ascend the red giant branch (RGB).  This
off-center flash is the first of a few successive helium shell
subflashes that remove the core electron degeneracy over 2 Myrs,
converting the object into a He burning star.  Though characterized by
Thomas over 40 years ago, this core flash phase has yet to be
observationally probed.  Using the Modules for Experiments in Stellar
Astrophysics ($\MESA$) code, we show that red giant asteroseismology
enabled by space-based photometry (i.e. {\it Kepler} and CoRoT) can
probe these stars during the flash. The rapid ($\lesssim 10^5 {\rm
  yr}$) contraction of the red giant envelope after the initiating
flash dramatically improves the coupling of the p-modes to the core
g-modes, making the detection of $\ell=1$ mixed modes
possible for these 2 Myrs. This duration implies that 1 in 35
stars near the red clump in the HR diagram will be in their core flash
phase.  During this time, the star has a g-mode period spacing of
$\Delta P_{\rm g}\approx 70-100\ {\rm s} $, lower than the $\Delta
P_{\rm g}\approx 250 \ {\rm s}$ of He burning stars in the red clump,
but higher than the RGB stars at the same luminosity. This places them
in an underpopulated part of the large frequency spacing ($\Delta
\nu$) vs.  $\Delta P_{\rm g}$ diagram that should ease their
identification amongst the thousands of observed red giants.
  
\end{abstract} 


\section{Introduction}

Sensitive space-based photometry from CoRoT \citep{baglin09} and
{\it Kepler} \citep{borucki09} has enabled detection and
characterization of radial and non-radial acoustic (i.e. p-mode) oscillations in
thousands of red giant stars \citep{deridder09,bedding10,huber10,mosser10,hekker11,miglio11}  
with frequencies centered at 
\begin{equation} 
\label{eq:numax}
\nu_{\rm max}\approx 31 \mu {\rm Hz} \left(M\over M_\odot\right)\left(10R_\odot\over R\right)^2\left(5777 \ {\rm K}\over T_{\rm eff}\right)^{1/2}, 
\end{equation} 
consistent with the scaling of the acoustic cutoff frequency \citep{brown91}. 
These modes generate photometric variability ranging from 5 to 1000 parts per
million \citep{mosser11,huber11} and have lifetimes
of at least  15  days \citep{huber10,baudin11}.  Their
excitation and damping (and thereby their resulting amplitudes;
\citet{kjeldsen95}) is  related to the presence of
vigorous convection within the star (see review by \citet{christen11}). 
  The majority of these oscillations are
p-modes with nearly evenly spaced frequencies at 
\begin{equation} 
\label{eq:delnu} 
\Delta \nu\approx4.27 \mu {\rm Hz} \left(M\over M_\odot\right)^{1/2}\left(10 R_\odot\over R\right)^{3/2}, 
\end{equation} 
a relation  used with equation (\ref{eq:numax}) to determine the
stellar mass, $M$, and radius $R$ from the measured $\nu_{\rm max}$,
effective temperature $T_{\rm eff}$ and $\Delta \nu$ \citep{hekker11}. 
The implied relation between $\nu_{\rm max}$ and $\Delta
\nu$ has also been confirmed observationally \citep{stello09,hekker09,bedding10, huber10, hekker11, mosser11,huber11}.

Space-based observations  \citep{bedding10, beck11,mosser11mixed} also enabled the detection of the 
angular degree $\ell=1$ mixed modes, which have p-mode
characteristics in the red giant envelope, but g-mode
characteristics in the helium core \citep{scuflaire74,osaki75,aizenman77,
  dziembowski01, christen04, dupret09, montalban10}.
Modes nearly evenly spaced in period, at $\Delta P_{\rm obs}$, around the $\ell=1$ p-modes were
identified by \citet{beck11} as a characteristic of the interior core g-modes 
allowing \citet{bedding11} to distinguish first ascent red giant branch (RGB) stars (i.e. those with degenerate helium cores) from red clump stars (i.e. those with non-degenerate He burning cores). This  separation in the $\Delta P_{\rm obs}-\Delta \nu$ diagram was also seen by CoRot \citep{mosser11mixed}, and is a powerful new tool for stellar population studies. 

We show here that this  new capability to probe the deep interior of a
red giant should allow for the identification of those $M \lesssim 2M_\odot$ 
stars undergoing the helium core flash. Known for more than 40 years as the defining event that ends the ascent of low mass stars up the RGB (and defines the tip of the RGB; \citet{salaris02}), this thermally unstable and off-center helium burning leads to a $t_f\approx 2$ Myr phase of successive He subflashes 
\citep{thomas67,ibenrenzini84,mocak08} that remove electron degeneracy and convert the He core to a stably burning non-degenerate object. However, there has been debate (see review by \citet{ibenrenzini84}) as to whether the initiating flash can become dynamical in some way, or rather remains hydrostatic. 

Using the $\MESA$ code \citep{paxton11},  we 
start in \S 2 by describing the  changes in the red giant envelope and helium core during the core flash. 
We work in the  Wentzel, Kramers, Brillouin (WKB) 
approximation in \S 3, summarizing 
the p-mode and g-mode properties during the helium core flash.  We close in 
\S 4 by discussing how stars in this phase of evolution can be differentiated from the more numerous  populations of RGB and clump stars.

\section{Stellar Evolution during the Core Flash Phase} 

\begin{figure}
\plotone{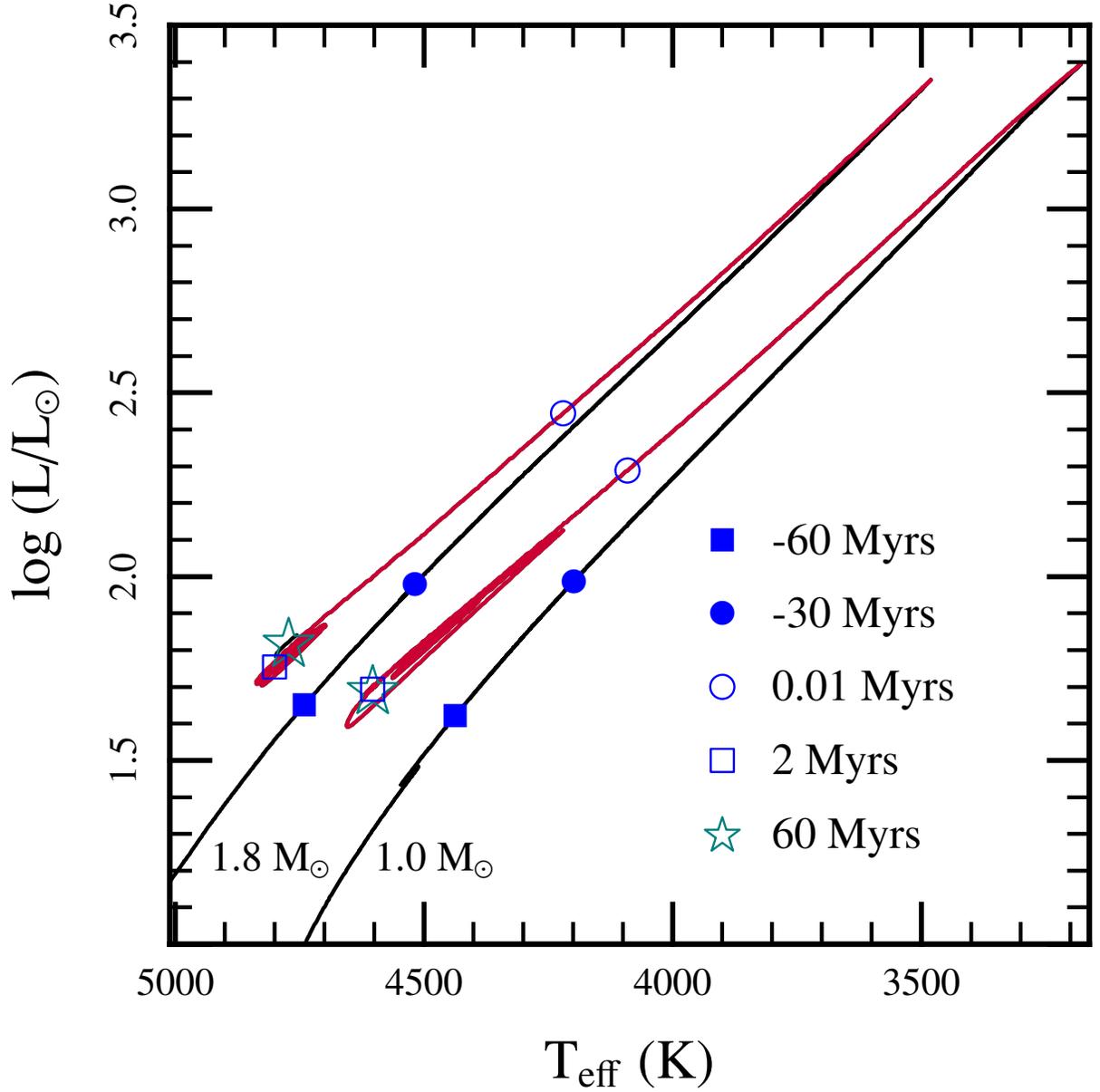} 
\caption{The evolution of a  solar metallicity 
$M=1.0M_\odot$ star and a $M=1.8M_\odot$ star before (black line), during  and after (red line) the  core flash phase.  The filled circles (squares) show where the star resided on the RGB  at 30 (60) Myrs prior to the initiating flash at $L_{\rm tip}=10^{3.4}L_\odot$. 
The open circles show  the  location  $0.01$ Myr after the initiating flash, whereas  the open blue square is 2 Myrs after the initiating flash, a time when the star reaches the red clump, where it still resides at 60 Myrs (green star). 
\label{fig:hrevol} }
\end{figure}

Evolved stars with $M\lesssim 2.0 M_\odot$ undergo a He core flash that ends their RGB 
evolution, and leads them to the red 
clump where they undergo core He burning for $\approx 70$ Myrs. The
flash starts as an off-center thermally unstable 
ignition of helium burning at a mass coordinate about half-way through the $M_c\approx 0.48M_\odot$ He core \citep{dominguez99,salaris02,serenelli05,serenellif05,mocak09, paxton11}, nearly  independent of metallicity and $M$.

 Using $\MESA$ version 3709, we have calculated this evolution for a range of $M$, all with $X=0.7$ and $Z=0.02$, treating  convection with the Schwarzshild criterion (i.e. no semiconvection, thermohaline mixing or convective overshoot) and  a mixing length parameter of $\alpha_{\rm MLT}=2.0$. We had no mass loss, diffusion, or rotation and we used the $\MESA$ ``basic" \citep{paxton11} nuclear network with rates from NACRE. The opacity was from OPAL \citep{iglesias96}, with the low temperature opacities taken 
from \citet{ferguson05} with 
metal ratios given by \citet{grevesse98}, as described in \citet{paxton11}.   We used the OPAL equation-of-state \citep{rogers02}, and HELM extensions \citep{timmes00} where needed. 

The evolutionary tracks in Figure \ref{fig:hrevol} for a $M=1.0M_\odot$ and $M=1.8M_\odot$ star are for the time before, during,  and after the core flash phase. 
The  He core expansion from the initiating flash leads to an adiabatic temperature drop  in the overlying H burning shell; quenching the burning. 
This loss of an energy source for the red
giant envelope triggers a Kelvin-Helmholtz (KH)  contraction from the tip of
the RGB on a rapid timescale \citep{thomas67},  reaching 
$L \approx 100-200  L_\odot$ in $10^4$ years (the open blue circles). 

\begin{figure}
\plotone{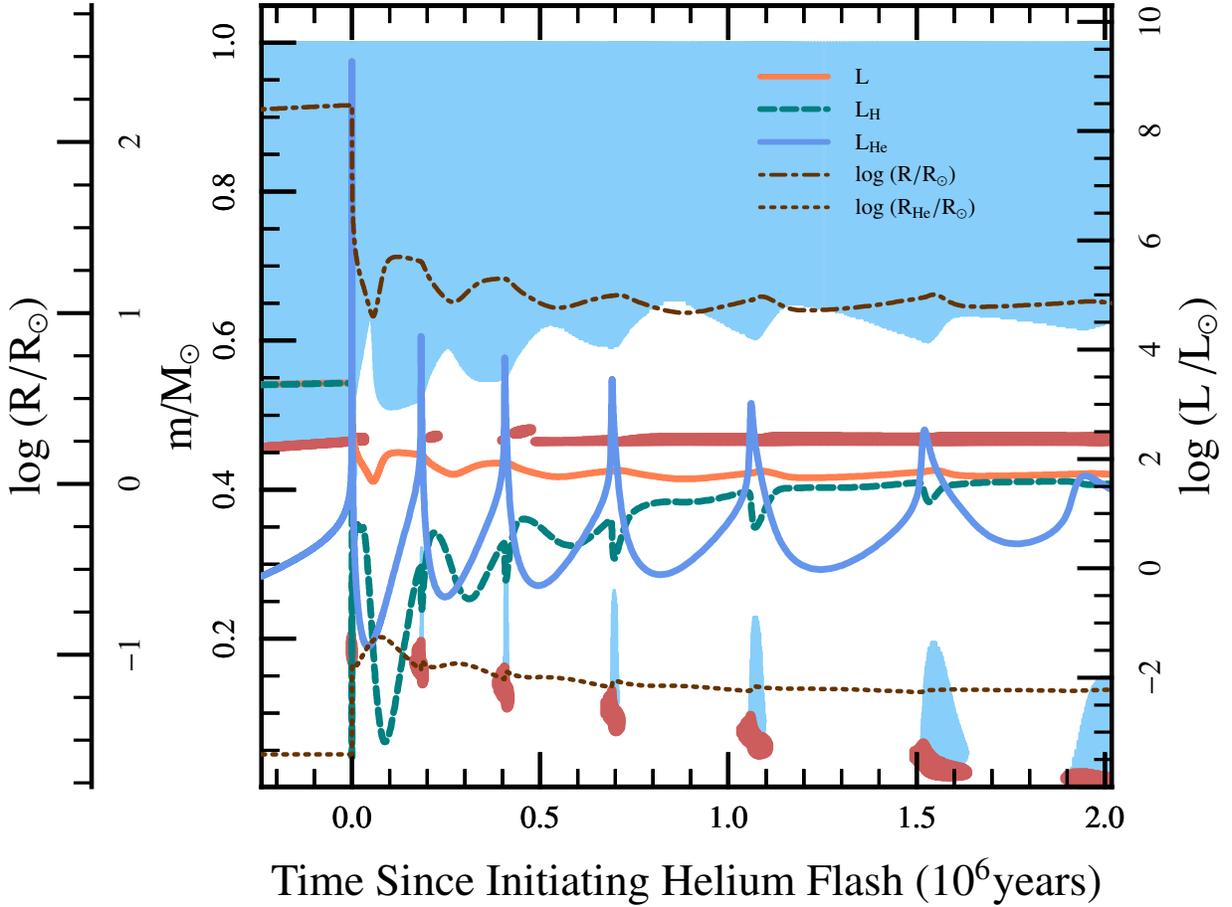} 
\caption{Properties of a $M=1.0M_\odot$ star for the 2 Myr core flash
  phase.  The blue shaded regions show where convection is present in
  mass coordinates, whereas the orange shaded regions are where
  nuclear burning yields $>1000 \ {\rm ergs \ gr^{-1} \ s^{-1}}$. The
  stellar luminosity is shown by the orange solid line, whereas the H
  (He) burning luminosities are shown by the green dashed (blue
  solid) lines. The stellar radius, $R$, is shown by the brown
  dot-dashed line, and the dotted brown line shows the He core radius.
\label{fig:structure} }

\end{figure}

\citet{thomas67} showed that the He core flash phase is governed by a series of shell subflashes that increase the entropy of the convective regions. This is seen in Figure \ref{fig:structure}, where five successive subflashes are seen to move inwards in mass coordinates. However, convective He burning is present for only about 10\% of the time. The 
$t_f\approx 2$ Myr  timescale is set by the need for thermal diffusion to act inwards between  subflashes, ending when the thermal wave reaches the core \citep{thomas67, serenelli05}, heating it at nearly constant pressure \citep{paxton11} to the condition needed for the convective He burning core phase of the red clump. We end Figure \ref{fig:structure} at that time. Even though the He burning luminosities (blue dashed lines)  get quite large ($L_{\rm He}\approx 10^{9.3}L_\odot$ in the initiating flash at $t=0$), \citet{shen09} have shown that the heating from He burning always occurs on a timescale much longer than the local dynamical time, an outcome confirmed in multi-dimensions by \citet{mocak08,mocak09}. The stellar response during the 2 Myr  core flash phase is shown in Figure \ref{fig:structure} and form the  basis for our asteroseismological work.
 
\section{Mode Properties during the  Core Flash Phase} 

\begin{figure}
\plotone{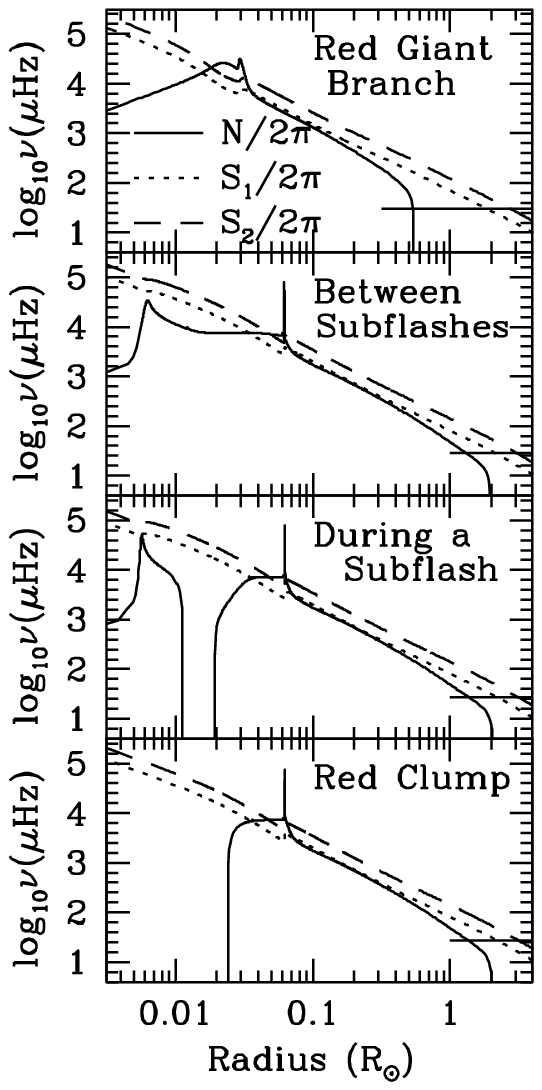} 
\caption{Propagation diagrams for a $M=1.0M_\odot$ star at four different stages of evolution (described in text), all when
$\Delta \nu=4 \mu$Hz. The Brunt-V\"ais\"ala frequency, $N/2\pi$ (solid lines), and Lamb frequencies $S_1/2\pi$ (dotted lines) and $S_2/2\pi$ (dashed lines) are shown. The thin horizontal line is at $\nu_{\rm max}\approx 28\mu$Hz. From top to bottom, these models have 
$L/L_\odot=41.6, 50.6, 51.1 \& 50.8 $, and $R/R_\odot=10.9, 11.2, 11.2 \& 11.2 $.
\label{fig:brunt} }
\end{figure}

The non-radial adiabatic mode structures for RGB and 
red clump stars have been studied, with focus on the 
$\ell=1$ mixed modes that reveal core properties
\citep{dziembowski01, christen04, dupret09, montalban10,christen11,christen11kep, jiang11, dimauro11}. 
These mixed modes have a p-mode quality in the outer parts of the star (where they are  excited by convection, and typically have $n_p\approx 10$  radial nodes), but penetrate into the stellar core as very high order ($n_g>100$) g-modes.  This justifies 
our initial exploration 
in the WKB limit \citep{unno89,aerts10,christen11},  where the local (at radius $r$) radial wavenumber, $k_r$,  is 
\begin{equation} 
k_r^2={1\over c_s^2 \omega^2}\left (\omega^2-N^2\right)\left(\omega^2-S_\ell^2\right), 
\label{eq:kr}
\end{equation}
where $\omega=2\pi \nu$ is the mode frequency, $c_s$ is the sound speed, 
$N^2$ is the Brunt-V\"ais\"ala frequency, and 
$S_\ell^2=c_s^2 \ell (\ell+1)/r^2$ is the Lamb frequency. Figure \ref{fig:brunt} shows the propagation 
diagrams (radial profiles of $N$, $S_1$ and $S_2$)
for four phases of the $M=1.0M_\odot$ model, all chosen when $\Delta \nu=4 \mu$Hz. These 
resulted in nearly the same values of $\nu_{\rm max}\approx 28 \mu$Hz (denoted by the horizontal line). 
From top to bottom, the panels are for $t=-60$  Myrs prior to the initiating flash (on the RGB), $t=1.4$ Myrs (after 
the initiating flash), $t=1.6$  Myrs (during a convective He burning subflash) and 
$t=2.0$ Myrs when the star is in the convective He core burning phase of the red clump. 

Modes supported by the acoustic response of the envelope (i.e. p-modes) propagate in the outer parts of the star  where $\omega^2>S_\ell^2$  and $\omega^2>N^2$. Since $N^2$ is either zero or very small there, we set 
$c_s^2 k_r^2=\omega^2-S_\ell^2$. The eigenfrequencies are then found 
by setting $\int k_r dr \approx n_p \pi $. In the extreme limit of $\omega^2 \gg S_\ell^2$, this simplifies to $n_p\pi=\omega \int dr/c_s$, where the integral extends over the outer parts of the star where $\omega^2>S_\ell^2$. Tradition is to write this as $\nu=n_p\Delta \nu$, where the ``large spacing'' $\Delta \nu$ is defined with the integral over the whole star,
\begin{equation} 
\Delta \nu^{-1}=2 \int_0^R  {dr \over c_s},
\label{eq:deltanucalc}
\end{equation} 
a reasonably accurate representation. 

  Modes supported by the internal buoyancy (i.e. g-modes) propagate in the stellar interior, 
where $\omega^2<N^2$ and $\omega^2<S_\ell^2$. As evident from Figure \ref{fig:brunt},  these inequalities are strong for frequencies 
$\sim \nu_{\rm max}$, yielding 
$k_r^2\approx \ell(\ell+1)N^2/r^2\omega^2$. The high radial order ($n_g\gg 1$)  g-modes then have 
$n_g\pi=(\ell(\ell+1))^{1/2}\int N d\ln r/\omega$, yielding
$\nu^{-1}=n_g\Delta P_g$, with $\Delta P_g$ being the period
spacing. For $\ell=1$, this is 
\begin{equation} 
\Delta P_g (\ell=1)={2^{1/2}\pi^2\over \int N d\ln r}. 
\label{eq:deltapg}
\end{equation} 
 At $\nu_{\rm max}$, $n_g\approx 1/(\nu_{\rm max}\Delta P_g)>100$, safely in the WKB limit. The integral $\int N d \ln r$ extends over the region where $\omega^2<N^2$ and $\omega^2<S_\ell^2$. 

\begin{figure}
\plotone{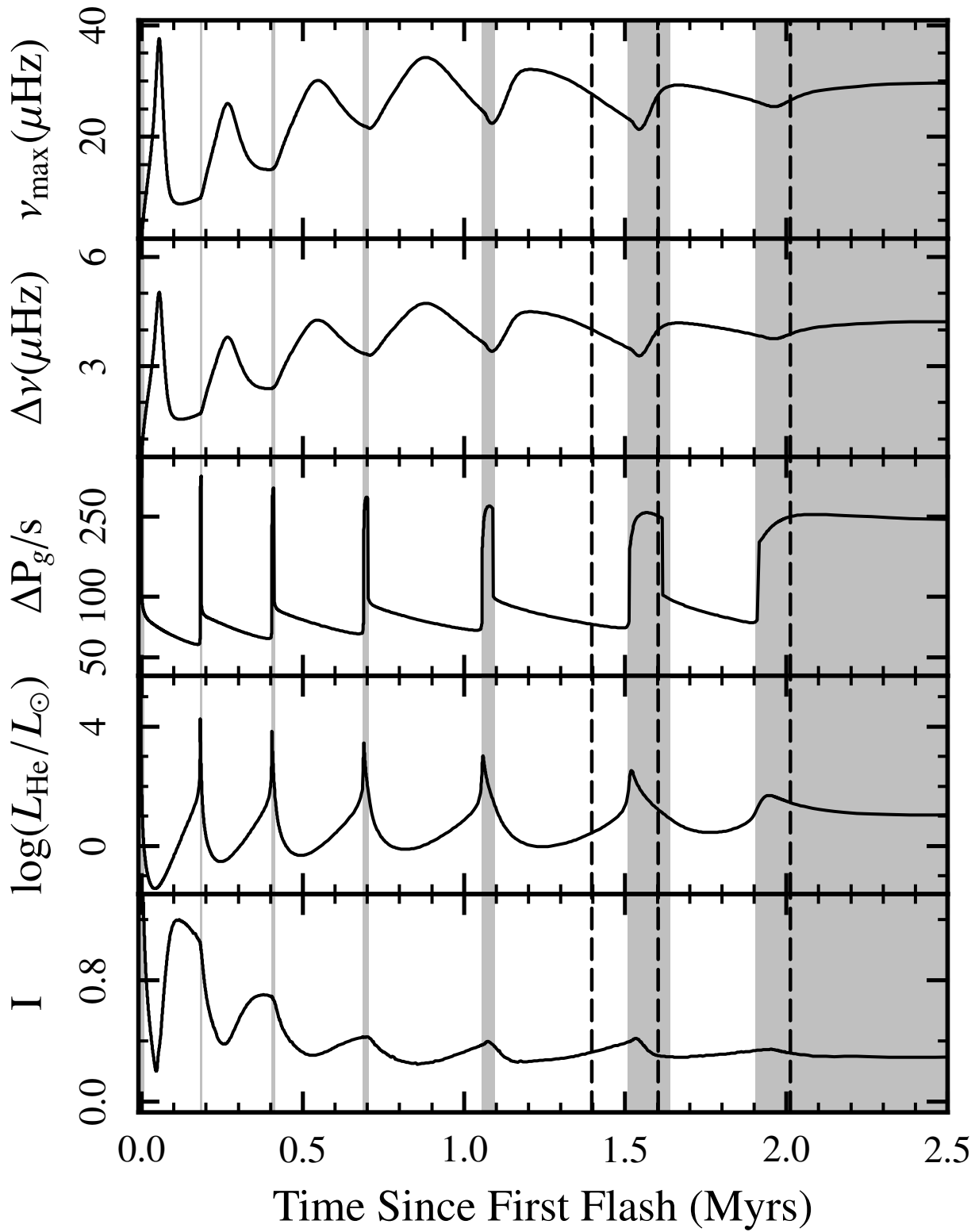} 
\caption{Seismic properties during the core flash, with vertical grey
regions showing subflashes. From top to 
bottom, the panels show $\nu_{\rm max}$  from equation (\ref{eq:numax}), $\Delta \nu$ from equation (\ref{eq:deltanucalc}), $ \Delta P_g$ from equation (\ref{eq:deltapg}), the Helium burning luminosity $L_{\rm He}$,  and the 
evanescent integral $I$. The vertical dashed lines denote the times 
for the bottom three panels of Figure \ref{fig:brunt}.
\label{fig:seismo} }
\end{figure}

Now consider how $\Delta P_g$ and $\Delta \nu$ evolve during the core
flash phase. The KH contraction after the initiating flash triggers a
rapid decrease to $R\approx 11R_\odot$, and the increase of $\Delta
\nu$ evident in the second panel of Figure
\ref{fig:seismo}. Thereafter, the value of $\Delta \nu$ simply
responds to the mild radius changes during subsequent subflashes. The
He core radius expansion and outer envelope contraction triggered by
the initiating He flash leads to an outer envelope structure during
the $t_f\approx 2$ Myr that is nearly the same as that on the red
clump. Indeed, the profiles of $N^2$ and $S_\ell^2$ in Figure
\ref{fig:brunt} for the bottom three panels (all after the initiating
flash) are nearly identical for $r>0.04R_\odot$.

The prime property that is changing over the 2 Myr is the He core, as
it undergoes additional subflashes that lift the degeneracy of the
deep interior. The $\ell=1$ g-mode period spacings for the models in
Figure \ref{fig:brunt} are $\Delta P_g=61.2$ s for the RGB, $\Delta
P_g=73$ s for the core flash phase model between subflashes (second
panel), and $\Delta P_g=255$ s for the red clump. During the
convective He burning shell subflashes (less than 10\% of the He core
flash phase, so $< 2\times 10^5$ years), an intervening evanescent
zone appears in the core (see third panel in Figure
\ref{fig:brunt}). Though coupling through this zone is possible (see
below), we calculated the period spacing assuming that only the outer
g-mode cavity is relevant, giving $\Delta P_g\approx 250-270$ s during
each sub-flash. As evident in the third panel of Figure
\ref{fig:seismo}, the existence of a degenerate core with $N^2>0$
during the most of the core flash period keeps $\Delta P_g<100$ s. The
fully convective He burning core of the red clump is what causes the
increase to $\Delta P_g=255$ s \citep{christen11} at the end of Figure
\ref{fig:seismo}.

 The $\ell=1$ mixed modes are more likely to be detected when the
 coupling through the outer evanescent region (the region in the star where
 $\omega^2>N^2$ and $\omega^2<S_\ell^2$ so that $k_r^2<0$) is
 strong. In the WKB limit, the ratio of the mode amplitudes (and
also location of the mode energy)  between
 the two turning points, the inner one at $r_1$ (where $\omega^2<N^2$
 first occurs) and the outer one at $r_2$ (where $\omega^2>S_\ell^2$
 first occurs) is largely determined by $I=\int_{r_1}^{r_2} k_r dr$,
 but also depends on $r_1$ and $r_2$ \citep{aerts10}.\footnote{In the
 plane-parallel limit (e.g. $r_2-r_1\ll r_2$), the mode amplitude
 ratio is $\propto \exp(\pm I)$.} The coupling of
 the $\ell=1$ modes is stronger than that of $\ell=2$ simply because
 the outer turning point is closer to the core for $\ell=1$
 \citep{christen04,dupret09}. The values of $I$ (calculated at
 $\nu=\nu_{\rm max}$) for the models in Figure \ref{fig:brunt} are
 $I=1.26$ for the RGB, and $I\approx 0.3$ for the other three models
 (which have nearly identical outer envelopes).  The bottom panel of
 Figure \ref{fig:seismo} shows that $I$ changes little over the 2 Myr,
 giving us confidence that the seismic probe of the core will be as
fruitful for the He core flash stars as the red clump. 

We have not fully investigated the mode structure during the
subflashes. However, a few things can be said. The first is that an
additional inner evanescent zone appears during subflashes (the region
at $0.0115 < r/R_\odot<0.0192$ in the third panel of Figure
\ref{fig:brunt}) splitting the g-mode cavity into two zones, each with
their own distinct period spacing. This adds an additional evanescent
integral $I_i\approx [\ell(\ell+1)]^{1/2}d\log r\approx 0.72$ that
must be accounted for in the observable mode structure. Due to that
extra penalty, we chose to display (in Figure \ref{fig:seismo}) the
period spacing during the subflashes as that of the outermost g-mode
cavity. If the inner cavity proves to be adequately coupled, these
extra modes would cause oscillations in the observed period spacing
diagram. That full analysis awaits our future efforts.

\begin{figure}
\plotone{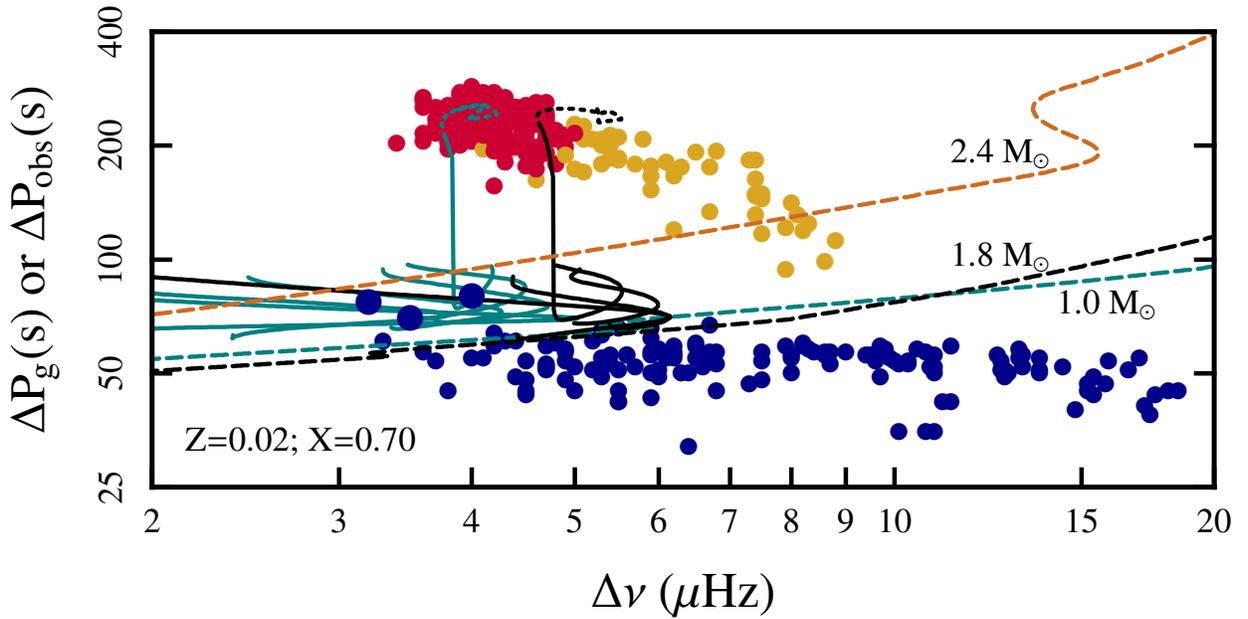} 
\caption{Evolution of $1.0M_\odot$ (green) and $1.8M_\odot$ (black) 
red giants through the helium core flash in $\Delta \nu-\Delta P_g$ space.  The RGB of a $2.4M_\odot$ star is  shown by an orange line. 
The dashed lines are the RGB phase, whereas the solid line is the core flash (with the rare excursions during subflashes omitted) and short-dashed the He burning stage. 
The colored data points are $\Delta P_{\rm obs}$ from  \citet{bedding11}. 
\label{fig:data} }
\end{figure}
  
\section{Detecting the  Core Flash Phase}

 Our results are best seen in relation to the observational work performed by \citet{bedding11} and \citet{mosser11mixed}. Their secure detection of mixed modes and measurement of their period spacing, $\Delta P_{\rm obs}$, 
 led to a clear distinction (see Figure \ref{fig:data}) of RGB stars (blue points) from the red clump stars (red points) of $M\lesssim 2M_\odot$ and the extended red clump (yellow points). The evolution of $\Delta P_g$ for 
our $1.0M_\odot$ (green lines) and a $1.8M_\odot$ (black lines) models  are shown in Figure \ref{fig:data} as a dashed line for the RGB, a solid line during the core flash and a short-dashed line  for the core He burning phase. 

\citet{bedding11} discussed the challenge of securely inferring
$\Delta P_g$ when only a few mixed modes are observed (see also
\citet{mosser11mixed,christen11kep}).  In particular, they found that the measured
spacing, $\Delta P_{\rm obs}$, would be less than $\Delta P_g$ by a
factor of $1.3-1.6$. Such a need for scaling is evident along the RGB,
and will be best resolved by longer duration {\it Kepler} data that
reveals additional mixed modes. This should prove possible given their
much longer expected lifetimes \citep{dupret09}.

Stars undergoing the core flash occupy a distinct region in Figure
\ref{fig:data}. The only other models that traverse this region are
$M>2.0M_\odot$ RGB stars, and potentially AGB stars, both reasonably
rare instances that will need to be observationally distinguished. The
simplest technique is to use the measured $\nu_{\rm max}$ values, as
$\nu_{\rm max}\approx 38$ Hz for the $2.2M_\odot$ and $2.4M_\odot$ RGB
models, whereas $\nu_{\rm max}<32$ Hz during the core flash for the
$1.0M_\odot$ model of Figure \ref{fig:seismo}. All objects with $70 
\ {\rm s}< \Delta P_{\rm g}< 200$ s and $2{\rm \mu Hz}<\Delta \nu < 6 {\rm \mu Hz}$ are worthy
of a serious analysis to see if they are explained as $\approx
1M_\odot$ core flash stars rather than $>2M_\odot$ RGB stars.  We
highlight three such candidates in Figure \ref{fig:data} by slightly
enlarging their blue points. Since $\Delta P_{\rm obs}$ is always less
than $\Delta P_g$, these points could only move upwards in this
diagram, placing them securely in the region of interest. There are
193 red clump stars in this figure, so we expect about $193/35\approx
5$ flashing stars, consistent with the few ``outliers" in this early
data set. \citet{miglio09} showed that the expected metallicity of
most of the CoRoT red giants is near solar, with a few as low as
one-tenth solar. Our $\MESA$ runs for one-tenth solar occupy a
comparable part of this diagram, though they go to slightly higher
values of $\Delta \nu$ during the flash due to reaching a lower
luminosity during KH contraction.

There is much to be learned from the asteroseismic discovery and
analysis of a star undergoing the core flash phase. It would be an
immediate confirmation of the $\sim 2$ Myr duration of the event;
eliminating the alternate possibilities of more dynamical
outcomes. Study and analysis of the g-modes may reveal the specific
stage of evolution for such a star in the 2 Myr phase.  Any possible
inferences regarding the rotation rate of the core during these 2 Myrs
would be especially valuable, as the short timescales may not allow
for perfectly rigid core rotation. Another path to detection of this
rare phase was highlighted by \citet{silva08}, who noted that RGB
stars with substantial mass loss will cross the RR Lyrae instability
strip on their way to the horizontal branch. Those stars would then
exhibit a detectable pulsation period change due to their secular
evolution.
 
Additional calculations are certainly needed, both to fully explore
  lower metallicities, as well as to see how additional mixing
  mechanisms (e.g. thermohaline) may impact the qualitative
  evolution. Fully consistent pulsation calculations are needed, both
  to confirm the WKB estimates here, but also to provide damping times
  and to explore the interesting periods of subflashes (only relevant
  for $\approx 2\times 10^5$ years) that may prove detectable if the
  sample size of red giants with measured $\Delta P_g$ gets large
  enough.

\acknowledgements

 We thank T. Bedding, T. Brown, J. Christensen-Dalsgaard and V. Silva Aguirre for discussions. This work was supported by the NSF through grants PHY 05-51164 and AST 11-09174 and by support for K.M and P.J.M. through the Worster Family Summer Fellowship fund.

\bibliographystyle{apj}

\end{document}